\documentclass[prl,twocolumn,showpacs,amsmath,amssymb,superscriptaddress]{revtex4-1}
\usepackage{dcolumn}
\usepackage{bm,graphicx}
\usepackage{etoolbox}
\usepackage{url}
\usepackage[colorlinks]{hyperref}
\usepackage{breakurl}
\usepackage{color}
\usepackage [latin1]{inputenc}

\begin{document}

\title{Landau level spectroscopy of PbSnSe topological crystalline insulator}

\author{Kristupas Kazimieras Tikui\v{s}is}
\email[]{krisviesoforas@gmail.com}
\affiliation{Charles University, Faculty of Mathematics and Physics, Institute of Physics, Ke Karlovu 5, 121 16 Prague 2, Czech Republic}
\affiliation{Institute of Physics, Academy of Science of the Czech Republic, Cukrovarnick\'a 10, Praha 6, CZ-16253, Czech Republic}

\author{Jan Wyzula}
\affiliation{Laboratoire National des Champs Magn\'etiques Intenses, Univ. Grenoble Alpes, CNRS-UPS-INSA-EMFL, 25 rue des Martyrs, B.P. 166, 38042 Grenoble Cedex 9, France}

\author{Luk\'a\v{s} Ohnoutek}
\affiliation{Charles University, Faculty of Mathematics and Physics, Institute of Physics, Ke Karlovu 5, 121 16 Prague 2, Czech Republic}
\affiliation{Centre for Photonics and Photonic Materials and Centre for Nanoscience and Nanotechnology, University of Bath, Bath BA2 7AY, U.K.}

\author{Petr~Cejpek}
\affiliation{Charles University, Department of Condensed Matter Physics, Ke Karlovu 5, 121 16 Prague 2, Czech Republic}

\author{Kl\'ara Uhl\'\i\v{r}ov\'a}
\affiliation{Charles University, Department of Condensed Matter Physics, Ke Karlovu 5, 121 16 Prague 2, Czech Republic}

\author{Michael~Hakl}
\affiliation{Laboratoire National des Champs Magn\'etiques Intenses, Univ. Grenoble Alpes, CNRS-UPS-INSA-EMFL, 25 rue des Martyrs, B.P. 166, 38042 Grenoble Cedex 9, France}

\author{Cl\'ement Faugeras}
\affiliation{Laboratoire National des Champs Magn\'etiques Intenses, Univ. Grenoble Alpes, CNRS-UPS-INSA-EMFL, 25 rue des Martyrs, B.P. 166, 38042 Grenoble Cedex 9, France}

\author{Karel V\'yborn\'y}
\affiliation{Institute of Physics, Academy of Science of the Czech Republic, Cukrovarnick\'a 10, Praha 6, CZ-16253, Czech Republic}

\author{Akihiro~Ishida}
\affiliation{Shizuoka University, Department of Electronics and Materials Science, 3-5-1 Johoku, Naka-ku, Hamamatsu 432-8561, Japan}

\author{Martin Veis}
\affiliation{Charles University, Faculty of Mathematics and Physics, Institute of Physics, Ke Karlovu 5, 121 16 Prague 2, Czech Republic}

\author{Milan Orlita}
\email[]{milan.orlita@lncmi.cnrs.fr}
\affiliation{Laboratoire National des Champs Magn\'etiques Intenses, Univ. Grenoble Alpes, CNRS-UPS-INSA-EMFL, 25 rue des Martyrs, B.P. 166, 38042 Grenoble Cedex 9, France}


\begin{abstract}
We report on an infrared magneto-spectroscopy study of Pb$_{1-x}$Sn$_x$Se,
a topological crystalline insulator. We have examined
a set of samples, all in the inverted regime of electronic bands, with the tin
composition varying from $x=0.2$ to 0.33. Our analysis shows
that the observed response, composed of a series of interband
inter-Landau level excitations, can be interpreted and modelled using
the relativistic-like Hamiltonian for three-dimensional massive Dirac electrons,
expanded to include diagonal quadratic terms that impose band
inversion. In our data, we have not found any clear signature of
massless electron states that are present on the surface of
Pb$_{1-x}$Sn$_x$Se crystals in the inverted regime. Reasons for this
unexpected result are discussed.
\end{abstract}

\maketitle

\section{Introduction}

Topological crystalline insulators (TCIs) are systems that belong to the broader class of materials with a band structure endowed non-trivial topological properties. TCIs are characterized by a well-defined bulk band gap and two-dimensional (2D) conical
bands on the surface. To a great extent, they resemble the better known topological insulators (TIs)~\cite{KanePHYSREVLETT2005,BernevigScience06,KonigScience07}. Nevertheless, the surface states of TCIs are not protected by time-reversal symmetry, but instead, by the mirror or rotational symmetry of the crystal lattice. The existence of TCIs was confirmed experimentally in ARPES studies of lead tin selenide~\cite{DziawaNATMAT2012}, thus verifying pending theoretical predictions~\cite{VolkovJETPLett85,PankratovSSC87,FuPHYSREVLETT2011,HsiehNATCOMMUN2012}.

Lead-tin selenide is a mixed crystal widely explored in the past. Intensive investigations of this compound started in sixties, see, e.g., \cite{MitchellPRB66,MartinezPRB73,MartinezPRB73II,MartinezPRB73III,MartinezPRB75,PreierAP79} or reviews \cite{Khokhlov02,Springholtz2014}, and were often motivated by potential applications in thermoelectrometry and optoelectronics -- in particular, by a possibility to use this material as an active medium in infrared lasers. THz/infrared magneto-spectroscopy played an important role in these pioneer studies by providing us with useful insights into the electronic band structure~\cite{CalawaPRL69,RamageJPC75,Bauer80,Bangepss82}.

 \begin{figure}[t]
\includegraphics[width=.44\textwidth]{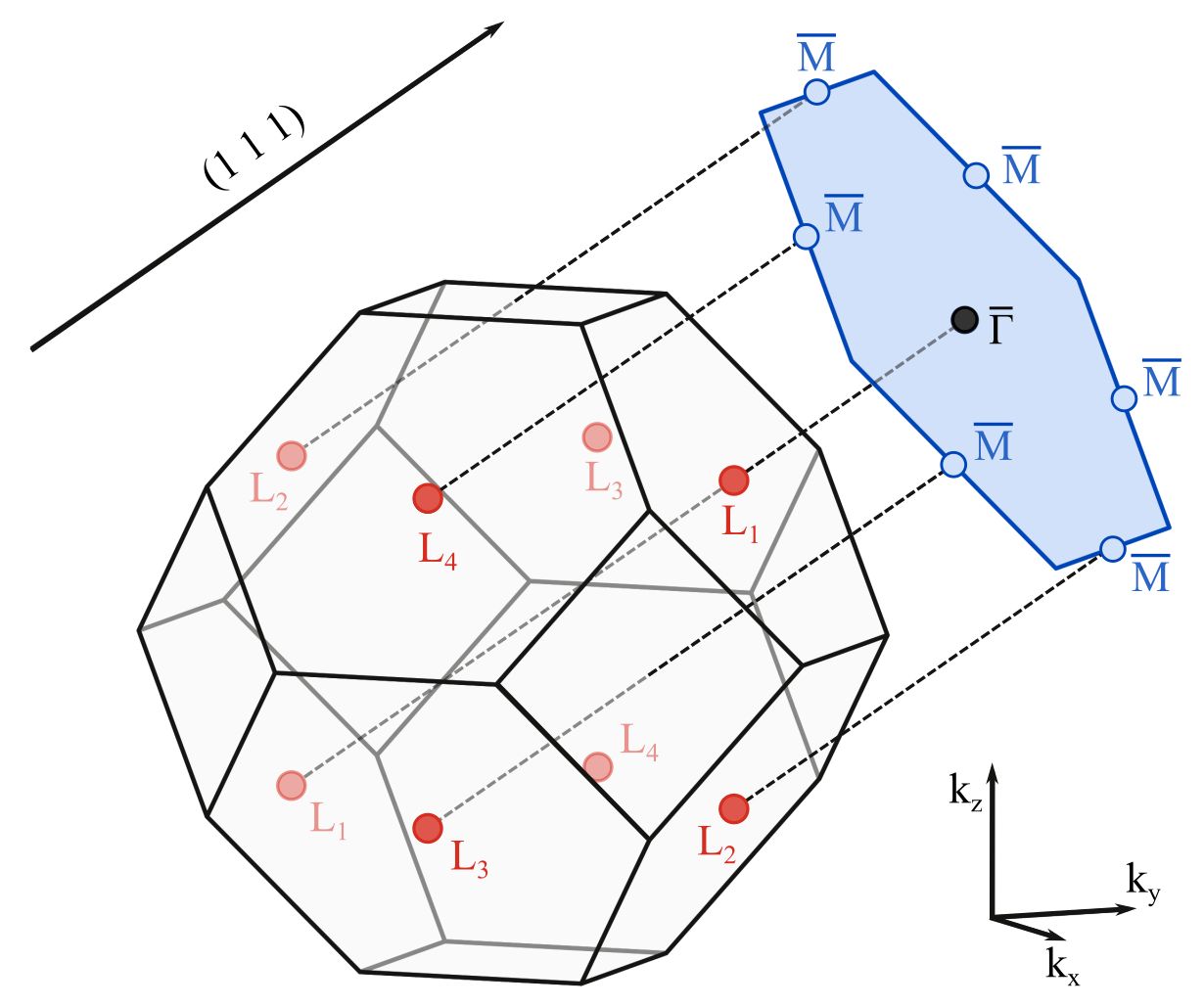}
\caption{Bulk Brillouin zone for a rock-salt lattice. Red points mark the positions of the four non-equivalent
$L$ points at the border of the zone. The blue hexagon shows the surface Brillouin zone in the (111) direction. The relativistic-like conical band appear at the $\bar{M}$ and $\bar{\Gamma}$ points, which are surface projections of the bulk $L$ points~\cite{LiuPHYSREVB2013}.}
\label{fig:BZ}
\end{figure}

The current consensus implies that Pb$_{1-x}$Sn$_x$Se is a narrow-gap semiconductor/semimetal. The fundamental energy band gap is direct and located at four non-equivalent $L$ points, the centers of hexagonal facets of the first Brillouin zone (Fig.~\ref{fig:BZ}; see also beginning of Sec.~IV for further details). The energy band gap drops with the increasing tin concentration $x$, and at liquid helium temperatures,
it closes at $x \approx0.15$~\cite{PreierAP79,WojekPRB14, KrizmanPRB18}. For higher tin compositions, the band gap re-opens but it becomes inverted and 2D conical bands appear on the surface~\cite{DziawaNATMAT2012,XuNATCOMMUN2012}. Such surface/interface states were predicted theoretically by Volkov and Pankratov~\cite{VolkovJETPLett85,PankratovSSC87}, and later on, classified using the
$Z_2$ invariant~\cite{FuPHYSREVLETT2011} -- this topological index takes on non-trivial value in a class of materials related to SnTe~\cite{HsiehNATCOMMUN2012}.
Nowadays, they continue to be a subject of intensive
theoretical~\cite{TchoumakovPRB17,AlspaughPRR20,MukherjeePRB19,LuEPL19} and experimental studies~\cite{AssafSR16,AssafPRL17,KrizmanPRB18,KrizmanPRB18II,KrizmanSPIE19}, including the search for their magneto-optical
signature~\cite{PhuphachongCRYSTALS2017,PhuphachongPhD17,WangNatureComm17}.

In this work, we explore a series of Pb$_{1-x}$Sn$_x$Se samples  using infrared magneto-spectroscopy and observe a relatively rich magneto-optical response due
to electronic excitations between pairs of bulk Landau levels. The analysis of the collected data allows us to extract the band structure parameters, and eventually, to get a quantified theoretical estimate for the magneto-optical response owing to surface electrons. Nevertheless,
our experiments performed in high magnetic fields -- strong enough to approach
the quantum limit of surface electrons and to probe their fundamental cyclotron mode -- do not show a clear signature for the magneto-optical
response originating from the surface states.

\begin{figure*}[t]
\includegraphics[width=.96\textwidth]{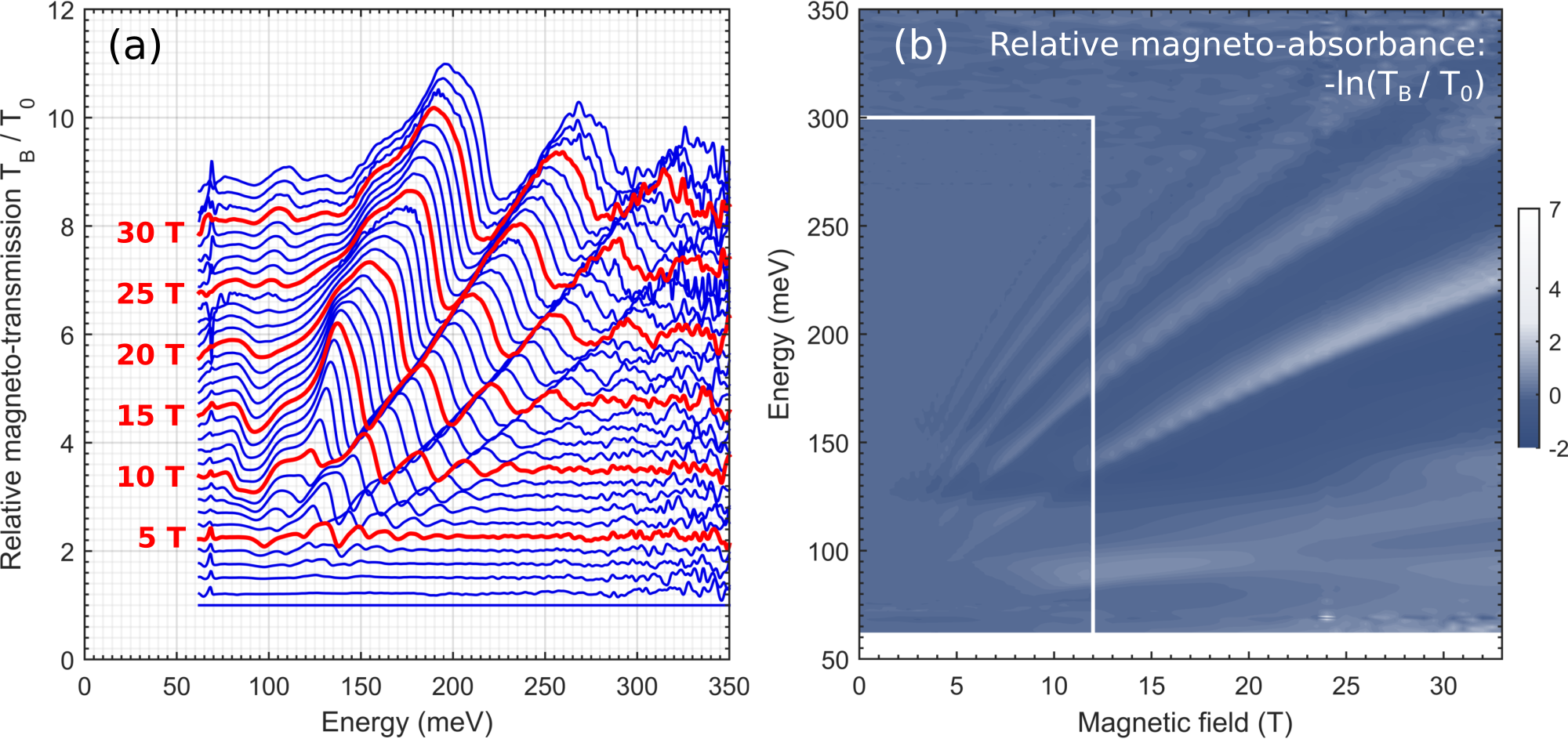}
\caption{Panel (a): The stacked-plot of relative magneto-transmission data, $T_B/T_0$, collected on sample C at $T=2$~K at magnetic fields up to $B=33$~T with the step of 1~T. For clarity, individual spectra were shifted vertically by ${B[T]\times\ 0.24}$. Panel (b): False color plot of relative magneto-absorbance, $A_B=-\ln(T_B/T_0)$,
measured on sample C. The white lines separate the areas of the data collected using a superconducting and resistive coil.}
\label{fig:exp_data}
\end{figure*}

\section{Preparation and characterization of PbSnSe samples}

A series of lead tin selenide (Pb$_{1-x}$Sn$_x$Se) samples (denoted here by letters A, B, C, D, E and F) was grown using the hot-wall-epitaxy method on a (111)-oriented BaF$_2$ substrate, with a weak doping by bismuth that ensures $n$-type conductivity. The thickness of epilayers ranged from 1 to 2.55~$\mu$m and the performed x-ray analysis indicated their (111)-orientation. The nominal tin content was chosen at $x=0.20$, 0.25 and 0.33. The composition of samples D and F was crosschecked using the energy-dispersive x-ray spectroscopy (EDS).The obtained results -- $x=0.22\pm0.02$ and $x=0.34\pm0.02$, respectively -- show a weak deviation from the nominal values. With the chosen tin contents, the samples are expected to have an inverted band structure at liquid-helium temperatures, but they are still away from the cross-over from the cubic (rock-salt) to orthorhombic phase that occurs
around $x\approx0.4$~\cite{DziawaNATMAT2012}.
All prepared samples were characterized using magneto-transport technique at room temperature which indicated $n$-type conductivity (Tab.~\ref{tab:sample_data}).

\begin{table}[h]
\caption{\label{tab:sample_data}
List of explored samples with the nominal Sn content $x$, carrier density $n$ and electron mobility $\mu$ deduced from room-temperature transport experiments:}
\begin{ruledtabular}
\begin{tabular}{lcccc}
Sample & $x$ & d [$\mu$m] & $\mu$ [cm$^2$/(V.s)] & $n$ [$10^{18}$ cm$^{-3}$] \\
\colrule
A & 0.20 & 0.8 & 720 & 10\\
B & 0.20 & 0.7 & 580 & 9.0 \\
C & 0.25 & 2.5 & 1200 & 1.4\\
D & 0.25 & 2.55 & 970 & 5.0\\
E & 0.25 & 2.5 & 1200 & 2.5\\
F & 0.33 & 1 & 350 & 11\\
\end{tabular}
\end{ruledtabular}
\end{table}

\section{Magneto-optical experiments}

The optical response of our PbSnSe samples was probed in the middle infrared spectral range using the magneto-transmission technique in the Faraday configuration, with the wave vector of light perpendicular to the sample surface. To measure magneto-transmission, unpolarized radiation of a globar was modulated in a Fourier-transform spectrometer and delivered to the sample using light-pipe optics.
Typically, several mm$^2$ of surface were exposed to the radiation which was then detected by a composite bolometer placed directly below the sample. The measured transmission spectra $T_B$ were corrected for the field-induced changes in the response of the bolometer. During experiments, the samples were kept in helium exchange gas at $T=2$~K and placed either in a superconducting (below 13~T) or resistive (above 13~T) coil.

\begin{figure*}[t]
\includegraphics[width=0.97\textwidth]{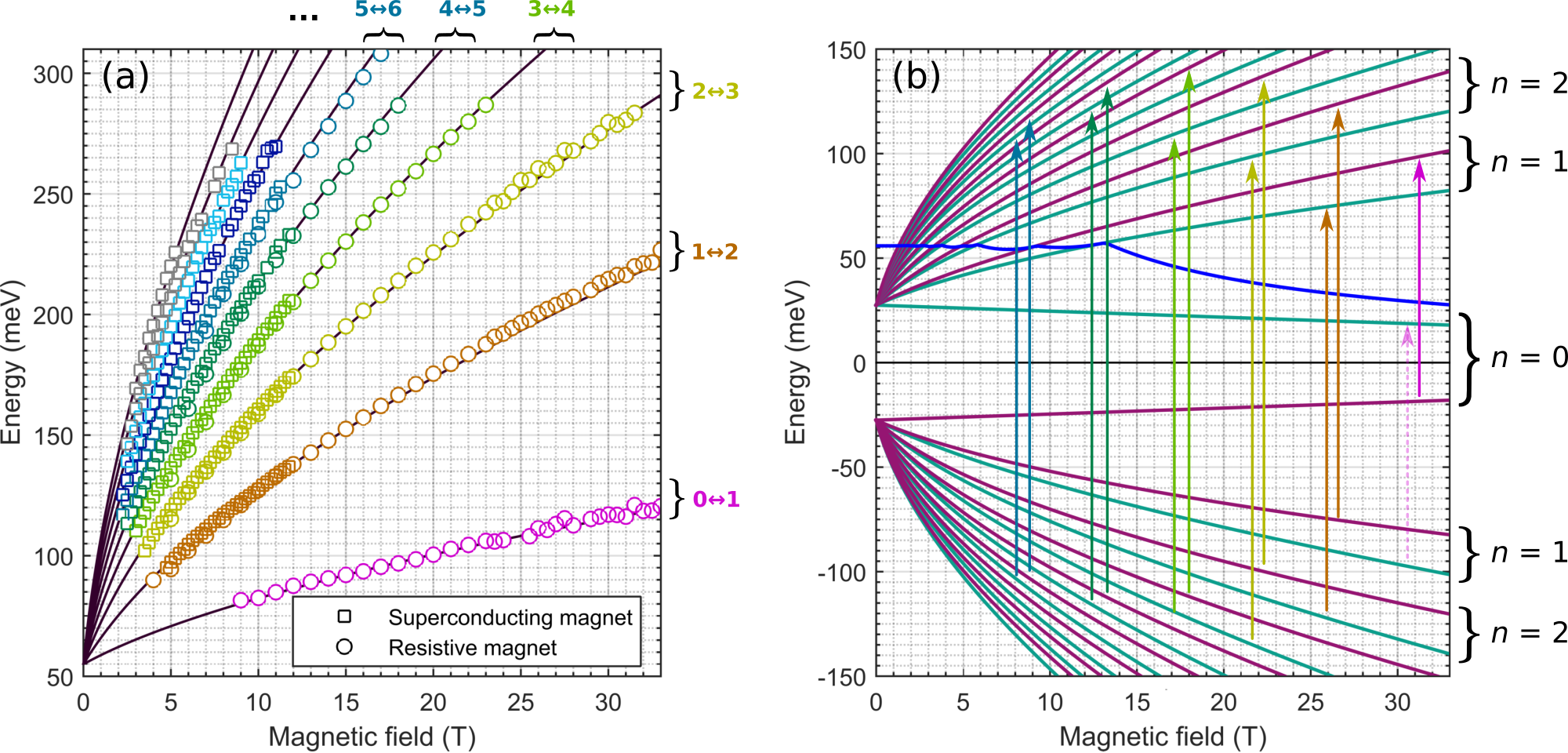}
\caption{
Panel (a):
Fan chart of inter-LL
excitations determined experimentally for sample C using a superconducting (squares)
and resistive coil (circles). The theoretically calculated energies of inter-LL excitations are plotted as solid lines for parameters deduced for sample C: $2\Delta = 55$~meV, $M=-20$ eV.{\AA}$^{2}$ and $v=4.4\times10^5$~m/s  (Tab.~\ref{tab:fit_params}).
Panel (b):
Landau level spectrum (at $q_z=0$) calculated for the parameters deduced for sample C. The levels plotted in dark cyan and violet colors correspond to the $\alpha=1$ and -1 LLs series, respectively. The blue line shows the calculated position of the Fermi energy assuming that the number of electrons
is independent of $B$. The vertical arrows show the interband inter-LL excitations active in the electric-dipole approximation.
The color coding in both panels has been
chosen to facilitate the identification of corresponding transitions.}
\label{fig:fan_diagram}
\end{figure*}

To illustrate the observed magneto-optical response, we have chosen the data collected
on sample C and plotted them in Fig.~\ref{fig:exp_data}.
In panel (a), we show a stack plot of relative magneto-transmission spectra, $T_B/T_0$, for selected values of the magnetic field. The false-color plot of relative magneto-absorbance, $A_B=-\ln(T_B/T_0)$ is presented in panel (b). At $B>2$~T,
the transmission of sample C becomes modulated by a series of excitations that follow a sub-linear dependence in $B$. These excitations may be straightforwardly identified as bulk inter-Landau level (inter-LL) transitions that promote electrons from the valence to conduction band.

To read out the positions of excitations, we have examined the differential magneto-transmission spectra $T_{B}/T_{B-\delta B}$, where $\delta B$ is the field step, and searched for the inflection points in regions with a negative slope. These can be, with a reasonable precision, associated with the
positions of excitations that shift relatively fast with the applied magnetic field. This procedure was implemented to suppress the impact of a pronounced Fabry-P\'erot interference pattern which appears
in the transmission spectra of epilayers at wavelengths comparable to their thicknesses and which are still partly visible in the $T_B/T_0$ spectra. The deduced fan chart of inter-LL transitions observed in the magneto-transmission of sample C is presented in Fig.~\ref{fig:fan_diagram}a. Up to ten individual transitions can be distinguished in the magneto-optical response of this particular sample.

\section{Magneto-optical response of bulk states -- theoretical model}

The parent compound of Pb$_{1-x}$Sn$_x$Se, rock salt structure lead selenide, is a semiconductor with a band gap around 0.15~eV at liquid helium temperatures~\cite{Springholtz2014,EkumaPRB12}. Close to the gap, both cation and anion atomic orbitals contribute significantly to the electronic states whereas Se (Pb) dominates in the single doubly-degenerate valence (conduction) band which transforms according to the $\Gamma_6^+$ ($\Gamma_6^-$) irreducible representation. An effective Hamiltonian can be derived using standard $\mathbf{k} \cdot \mathbf{p}$ perturbation theory, applied to
the first order and to a finite number of bands, to describe the
electronic states around the $L$ point of the Brillouin
zone~\cite{MitchellPRB66}: $\mathbf{q} = (q_x, q_y, q_z) =
\mathbf{k}- \mathbf{k}_L$, being the small parameter.

With a full electron-hole symmetry, which holds up to almost 1~eV away from the gap in PbSe~\cite{EkumaPRB12}, we therefore assume that the low-energy excitations in the bulk of Pb$_{1-x}$Sn$_x$Se can be described by
\begin{equation}
\label{eq:hamiltonianB0}
\hat{H_0}=\left( \begin{array}{cccc}
\Delta + M q^2  & \hbar v q_{+} & 0 & -\hbar v q_z \\
\hbar v q_{-} & -\Delta - M q^2 & \hbar v q_z & 0 \\
0 & \hbar v q_z & \Delta + M k^2 & \hbar v k_{-}\\
-\hbar v q_z & 0 &  \hbar v q_{+} & -\Delta - M q^2 \\
\end{array} \right),
\end{equation}
where $q_{\pm}=q_x\pm i q_y$ and $v$, $\Delta$ and $M$ are band structure parameters that depend on $x$. This Hamiltonian implies an isotropic dispersion for both conduction and valence bands ($\beta= 1\quad\mathrm{and} -1$, respectively):
\begin{equation}
\label{dispersion}
E_\beta(\mathbf{q})=\beta\sqrt{\left(\Delta + M q^2\right)^2+\hbar^2 v^2 q^2}
\end{equation}
with both dispersion branches doubly degenerate for spin.

The Hamiltonian \eqref{eq:hamiltonianB0} is equivalent to the three-dimensional Dirac Hamiltonian for massive electrons, where the velocity parameter $v$ can
be viewed as an effective speed of light. It is expanded to include dispersive diagonal
elements, $Mq^2$, which account for the impact of remote bands as well as the band inversion. For a relatively small inversion parameter, $|M|<\hbar^2v^2/\Delta$, the energy band gap of $2\Delta$ appears at $\mathbf{q}=0$. Notably, the Hamiltonian closely resembles the one proposed by Liu~et~al.~\cite{ZhangNaturePhys09,LiuPHYSREVB2010} for the electronic states around the $\Gamma$ point in the Bi$_2$Se$_3$-family of topological insulators.

The applied magnetic field ($B\,\|\,z$) transforms each electronic band into Landau levels. When the motion of electrons along the applied magnetic field is neglected ($q_z=0$), the LL spectrum reads:
\begin{equation}
E^n_{\alpha,\beta}\Bigr\rvert_{q_z=0}=\alpha \frac{M}{l_B^2}+\beta \sqrt{\left(\Delta+2\frac{M}{l_B^2}n\right)^2+2\frac{v^2 \hbar^2}{l_B^2}n},
\label{eq:Landau_bands}
\end{equation}
where $n= 1,2,3\ldots$ and $l_B=\sqrt{\hbar/eB}$ is the magnetic length. This spectrum describes LLs in the conduction and valence bands ($\beta=1$ and -1, respectively) as well as their splitting with respect to spin ($\alpha=\pm 1$). In addition, there exists a pair of $n=0$ LLs which are spin-polarized and which disperse linearly with $B$:
$E^0_{1,1}=\Delta+M/l_B^2$ and $E^0_{-1,-1}=-\Delta-M/l_B^2$. In the inverted regime, $M<0 $, these so-called zero-mode LLs (anti)cross each other, which is behaviour typical of all topological (crystalline) insulators~\cite{KonigScience07,OrlitaPRB11,AssafPRL17,KrizmanPRB18II}.

Knowing the LL spectrum and the corresponding eigenstates, one may proceed with the calculation of the dynamical magneto-conductivity that encodes the response in magneto-optical experiments. In the linear response theory, the dynamic conductivity -- for the right (+) and left (-) circularly polarized light -- is given by:
\begin{equation}
\sigma^{\pm}(\omega,B)=2\frac{iG_0 \zeta}{l_B^2 \omega}\sum_{n,m,q_z}\frac{(f_m-f_n)|\left<m|\hat{v}_{\pm}|n\right>|^2}{E^n-E^m-\hbar\omega+i\gamma},
\label{Conductivity}
\end{equation}
where $G_0=e^2/(2\pi\hbar)$ is the quantum of conductance, $\gamma$ is the broadening
parameter and $\zeta$ stands for the valley degeneracy ($\zeta=4$, see Fig.~\ref{fig:BZ}).
The occupation of LLs is given by the Fermi-Dirac distribution $f$ and the indices $m$ and $n$ run over all available initial and final LLs, including states with non-zero momenta $q_z$. The velocity operators $\hat{v}_{\pm}=\hat{v}_x\pm i \hat{v}_y$ are defined as $\hbar\hat{v}_{x}=\partial\hat{H}/\partial q_x$ and $\hbar \hat{v}_{y}=\partial\hat{H}/\partial q_y$, respectively,
and rewritten in the representation of ladder operators
$a^+=l_B(q_x-iq_y)$.

The matrix elements $\left<m|\hat{v}_{\pm}|n\right>$ contain the selection rules for electric-dipole excitations between various pairs of LLs. The dominant contribution to the magneto-optical response arises from the states with vanishing momenta $q_z$ for which those rules have a particularly simple form. The spin parameter $\alpha$ is always conserved and, for photon energies not exceeding band gap
considerably, only two series appear: the $\alpha=1$ series with the selection rule $n\rightarrow n-1$, and the $\alpha=-1$ series with $n\rightarrow n+1$. These are active for left and right circularly polarized radiation, respectively (see, e.g., Refs.~\cite{OrlitaPHYSREVLETT2015,LyJPCM16}).

\section{Magneto-optical response of bulk states -- discussion}

Let us now confront our experimental data with predictions of the sketched theoretical model.
In the first step, we compare the positions of inter-LL excitations deduced from our relative magneto-transmission/absorbance data with theoretical expectations.
Such a comparison is illustrated using the experimental data collected on sample C (Fig.~\ref{fig:fan_diagram}a). Individual observed transitions were interpreted based on the response in high magnetic fields when bulk electrons supposedly reach their quantum limit. The interband transition with the lowest energy may then be assigned to the fundamental electric-dipole-active interband mode which promotes electrons from the hole-like $n=0$ level to the spin-up $n=1$ level (Fig.~\ref{fig:fan_diagram}b).

\begin{figure}[h]
\includegraphics[width=.36\textwidth]{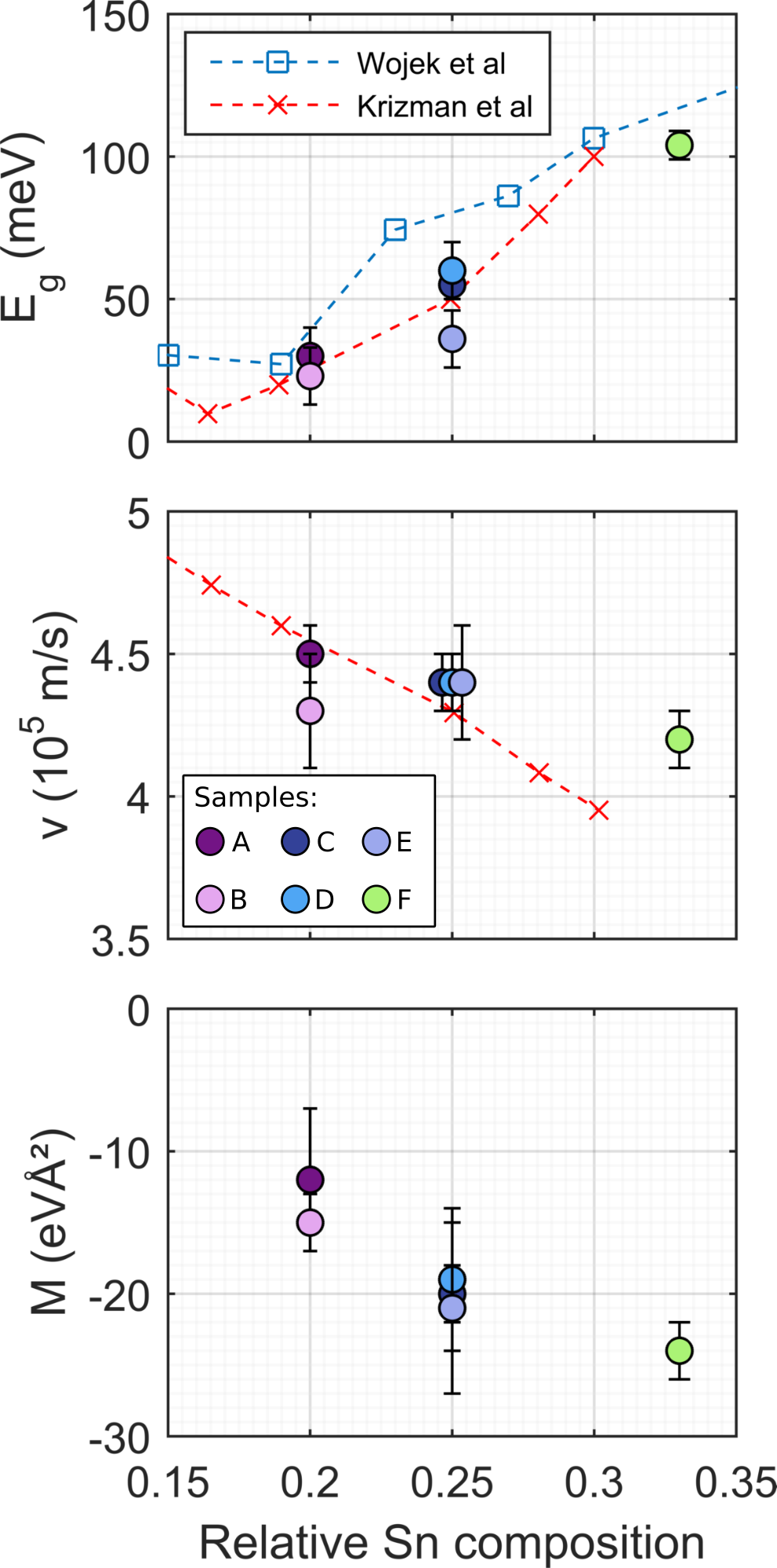}
\caption{Band structure parameters extracted from the magneto-optical data as a function of the nominal tin concentration: (a) energy band gap $E_g=2\Delta$, (b) velocity parameter $v$ and (c) inversion parameter $M$. The symbols connected by dashed lines indicate the band gap and velocity parameter values extracted in Refs.~\cite{WojekPRB14} and \cite{KrizmanPRB18}.}
\label{fig:fitted_parameters}
\end{figure}

Very good agreement between our simple theory and experimental data presented in Fig.~\ref{fig:fan_diagram}a corroborates the validity of the model. The positions as well as the field dependence of all observed transitions are reproduced by using only three adjustable parameters $\Delta, M$ and $v$. Notably, our model accounts for the
response of both longitudinal and oblique valleys (L$_1$ versus L$_{2,3,4}$ in Fig.~\ref{fig:BZ}).  Therefore, no anisotropy of the velocity parameter $v$ had to be considered in order to explain the magneto-optical data collected on sample C (Figs.~\ref{fig:exp_data} and \ref{fig:fan_diagram}) neither on other explored samples. This conclusion agrees with a relatively low anisotropy of
the electronic bands at the $L$ point in PbSe (expressed, e.g., in terms of the anisotropic effective masses  ~\cite{Bauer80,Springholtz2014}) which further decreases upon incorporation of tin~\cite{KrizmanPRB18}.
This is in contrast to some other lead-salt compounds, such as PbSnTe, in which the anisotropy is profound~\cite{MuraseSaM85,AssafSR16}, thus giving rise to well-resolved splitting of interband inter-LL resonances due to oblique and longitudinal valleys.

The deduced values of parameters for all six explored samples are listed in Tab.~\ref{tab:fit_params} and visualized in Fig.~\ref{fig:fitted_parameters}. We observe that the energy band gap ($E_g=2\Delta$) increases roughly linearly with tin content. The same behaviour is observed for the absolute value of the inversion parameter $|M|$. The difference in the extracted band gap for samples with presumably the same tin concentration points towards a certain variation of $x$ as compared to the declared nominal values, in line with the EDS results mentioned before. The deduced velocity parameters show a weak tendency to decrease with $x$. Such behaviour is perfectly in line with previous studies~\cite{Bauer80,AssafNPJQUANTMATER2017,KrizmanPRB18}.


\begin{table}[h]
\caption{\label{tab:fit_params}
The energy band gap $2\Delta$, velocity $v$ and inversion parameter $M$ extracted from the magneto-optical response of the samples A, B, C, D, E and F.}
\begin{ruledtabular}
\begin{tabular}{lcccc}
Sample & Sn content & $2\Delta$ (meV) & $v$ (10$^5\,$m.s$^{-1}$) & $M$ (eV\AA$^2$) \\
\colrule
A & 0.20 & 30 $\pm$ 10   & $ 4.5 \pm 0.1  $ & -12 $\pm$ 5 \\
B & 0.20 & 23 $\pm$ 10   & $ 4.3 \pm 0.2  $ & -15 $\pm$ 2 \\
C & 0.25 & 55 $\pm$ 5   & $ 4.4 \pm 0.1  $ & -20 $\pm$ 2 \\
D & 0.25 & 60 $\pm$ 10 & $ 4.4 \pm 0.1  $ & -19 $\pm$ 5 \\
E & 0.25 & 36 $\pm$ 10 & $ 4.4 \pm 0.2 $ & -21 $\pm$ 6 \\
F & 0.33 & 104 $\pm$ 5 & $ 4.2 \pm 0.1 $ & -24 $\pm$ 2 \\
\end{tabular}
\end{ruledtabular}
\end{table}

Notably, the characteristic shape of the conduction and valence bands described by the Hamiltonian (\ref{eq:hamiltonianB0}) can be assessed using the parameter $\eta=4|M|\Delta/(\hbar^2 v^2)$. For $\eta=1$, the electronic bands described by Eq.~\ref{dispersion} are strictly parabolic: $E_\pm(\mathbf{q})=\pm(\Delta+|M|q^2)$ and the LL spectrum as well as the whole magneto-optical response then scale linearly with $B$. This occurs, for instance, at the $\Gamma$ point of the well-known Bi$_2$Se$_3$ topological insulator. For all PbSnSe samples investigated in this work, we obtained $\eta<1$. This implies sub-parabolic profiles of bands, and consistently with our experiments, magneto-optical excitations following a sub-linear dependence in $B$ (Figs.~\ref{fig:exp_data} and \ref{fig:fan_diagram}). An extrapolation of band-structure parameters as a function of the tin content (see Figs.~\ref{fig:fitted_parameters}a-c) suggests that the condition $\eta = 1$ might be in Pb$_{1-x}$Sn$_x$Se achieved at  concentrations slightly above $x=0.35$.

\begin{figure}
\includegraphics[width=1.0\columnwidth]{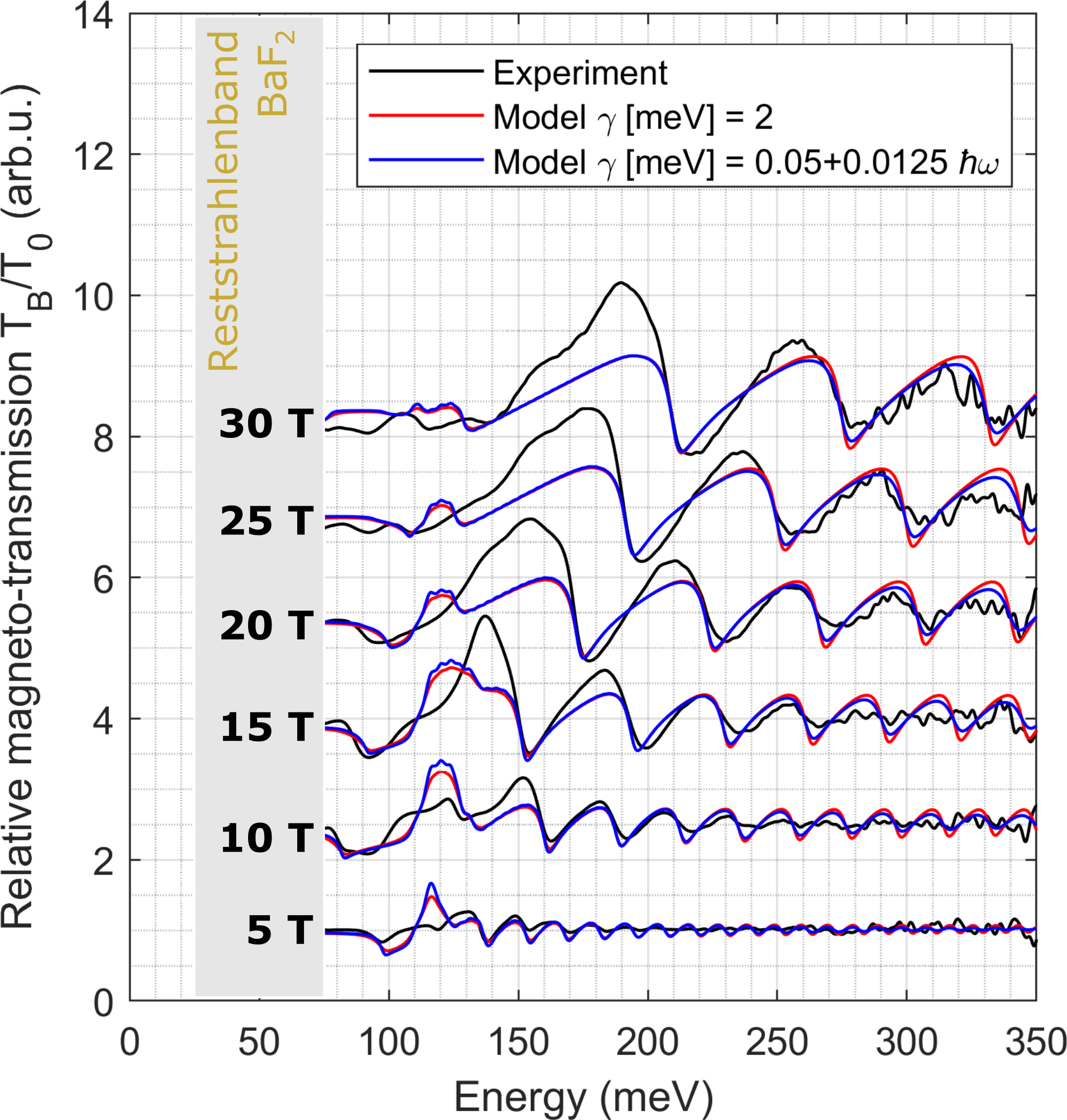}
\caption{Relative magneto-transmission spectra $T_B/T_0$ at selected values of the applied magnetic field compared to those calculated using the linear response theory~(\ref{Conductivity}). To calculate $T_B$ and $T_0$, the Landau levels up to the index 20 and 200 were included, respectively. The contribution of excitation at nonzero momenta were taken into account up to $q_z=1.25$~nm$^{-1}$. Two cases have been considered to account for the disorder in the system: $\gamma=2$~meV (red curve) and $\gamma=0.05$~meV+0.0125$\hbar\omega$~[meV] (blue curve).}
\label{fig:exp_theory_comparison}
\end{figure}

In the second step, we compare the experimentally determined relative magneto-transmission spectra, $T_B/T_0$, with those calculated theoretically, see Fig.~\ref{fig:exp_theory_comparison}. To obtain the theoretical spectra, we have numerically diagonalized the full Hamiltonian (\ref{eq:hamiltonianB0}) in which the magnetic field was introduced using the standard Peierls substitution. The calculated eigenstates (Landau levels/bands) and the corresponding spectrum were then used to evaluate the optical conductivity (\ref{Conductivity}). Afterwards, magneto-transmission spectra $T^{\pm}_B$
of the slabs with given thicknesses $d$ (Tab.~\ref{tab:sample_data}) were calculated for both circular polarizations of light~\cite{PalikRPP70}. The background dielectric constant, $\epsilon=20$, was estimated from the Fabry-P\'erot interference pattern (cf. Ref.~\cite{AnandPRB14}). The average transmission spectrum $T_B=(T^+_B+T^-_B)/2$ was then normalized by $T_0$, approximated by $T_{B}$ calculated at
the magnetic field below the onset of the LL quantization in the theoretical and also experimental response ($B=1$~T).

To reproduce the measured  magneto-transmission spectra, the  broadening parameter $\gamma$ had to be adjusted.
The simplest choice -- a parameter independent of the energy and magnetic field -- led to acceptable results (red curves in Fig.~\ref{fig:exp_theory_comparison}). The used value of $\gamma=2$~meV is consistent with the onset of the LL quantization observed in the magneto-optical data ($B\approx 2$~T).
The agreement is further improved by taking an effective value
of $\gamma$ that increases linearly with the photon energy: $\gamma=\gamma_0+\xi\hbar\omega$, with $\gamma_0 = 0.05$~meV and $\xi=0.0125$. This latter choice resembles the empirical
rule deduced for interband inter-LL excitations in graphene~\cite{OrlitaPRL11,NedoliukNatureNano19}. Notably, for relatively weak disorder, the width of excitations
is not entirely determined by the $\gamma$ parameter. The pronounced
high-energy tail of interband inter-LL excitations appears due to excitations at non-zero $q_z$ momenta. Moreover,
even though the electron-hole asymmetry was not directly evidenced
in our data, it may still influence the width of transitions. This latter contribution would be roughly equal to the difference between the cyclotron energies of electrons and holes.

The carrier concentration, or alternatively, the Fermi energy is another tunable parameter in our model. In our calculations, we considered the electron density to be constant as a function of $B$. The electron densities, tuned to achieve best possible agreement with the magneto-optical data, were roughly by a factor of two smaller as compared to results of room-temperature Hall experiments (Tab.~\ref{tab:sample_data}). The sample F was the only exception -- a carrier density almost one order of magnitude smaller than the one deduced from room-temperature characterisation had to be considered to reproduce our data. To adjust the Fermi energy, the appearance of the lowest bulk inter-band inter-LL excitation (0~$\rightarrow$~1) was taken as a main reference point.

The comparison of the experimental and theoretical  magneto-transmission spectra in Fig.~\ref{fig:exp_theory_comparison} indicates that fairly good agreement is achieved at higher photon energies, both for the shape and intensity of excitations. At lower energies, a certain deviation may be observed in the range corresponding to excitation of electrons to the vicinity of the Fermi energy. The relatively pronounced maximum in the theoretical $T_B/T_0$ spectra (around 120~meV in Fig.~\ref{fig:exp_theory_comparison}), is not observed experimentally. It appears in the calculated spectra as a result of $B$-induced splitting of the interband absorption edge for opposite circular polarisation of the incident light. In our experimental data, this feature seems to be smeared out due to the inhomogeneity of the electron density (fluctuation of the Fermi energy) across the explored sample.

\begin{figure*}[t]
\includegraphics[width=0.95\textwidth]{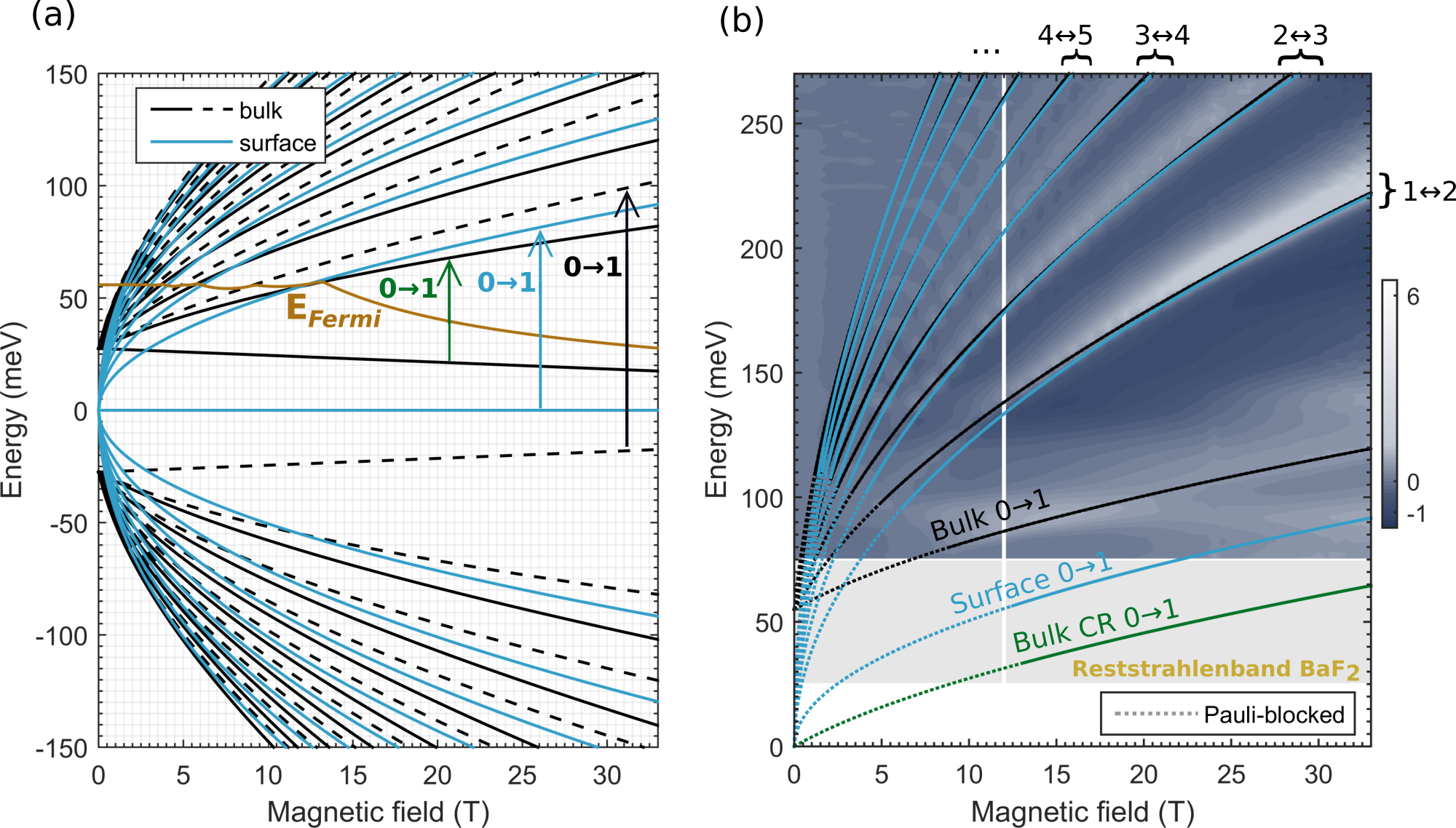}
\caption{Part (a): Landau level spectrum of surface and bulk states and the expected Fermi energy in sample C as a function of $B$. The dashed and solid lines used for the LLs reflect the spin parameter $\alpha = -1$ and 1, respectively.
The vertical arrows show three inter-LL excitations with lowest energies expected in the spectrum: the bulk interband 0~$\rightarrow$~1 transition, and fundamental CR resonances of bulk and surface electrons. Part (b): False-color plot of relative magneto-absorbance measured on sample C compared to the theoretically expected positions of bulk (black and green) and surface (dark cyan) inter-LL resonances. The dotted lines correspond to the values of the magnetic field at which the transitions are not active due to occupation of the final-state LL (Pauli blocking). The vertical white line separates the data measured at low magnetic fields using a superconducting coil from the high-field data obtained with a resistive coil.}
\label{fig:surface_states}
\end{figure*}

\section{Magneto-optical response of surface states}

Let us now focus on possible signatures of surface states in our magneto-optical data. In contrast to the (001)-oriented surface of PbSnSe samples~\cite{OkadaScience13}, the topologically non-trivial states on the (111) surface have, according to the existing theoretical models, a relatively simple structure~\cite{LiuPHYSREVB2013,HsiehNATCOMMUN2012}. Four conical bands are expected at the $\overline{M}$ and  $\overline{\Gamma}$ points which are projections of the bulk $L$ points where the electronic bands are inverted (Fig.~\ref{fig:BZ}).

The particular shape of conical bands -- their slope and anisotropy -- is given by the properties of the bulk velocity parameter. As discussed above, we have found no differences in the magneto-optical response of the longitudinal and oblique valleys. Therefore, we consider the velocity parameter to be isotropic in our PbSnSe samples in the explored range of tin content. In this approximation, both the surface and the interface of our PbSnSe epilayers host four identical conical bands with the slope defined by the bulk velocity parameter $v$ (Tab.~\ref{tab:fit_params}).

When a strong enough magnetic field is applied, 2D cones split into a series of non-equidistantly spaced LLs:
\begin{equation}
E_{\mathrm{2D}}^n= \pm v \sqrt{2 e\hbar  n B}.
\end{equation}
The magneto-optical response of systems with such a LL spectrum has been widely studied, both theoretically and experimentally, on graphene and other systems~\cite{SadowskiPRL06,JiangPRL07,HenriksenPRL10,OrlitaPRL11,SchafgansPRB12,BordacsPRL13,RussellPRL18,NedoliukNatureNano19,PackPRX20}. The response is dominated by a series of interband and intraband inter-LL excitations that follow the standard electric-dipole selection rules in isotropic systems: $n\rightarrow n \pm 1$.

When the quantum limit of electrons in a conical band is approached, the fundamental cyclotron resonance (CR) mode emerges in the optical response. This mode comprises excitations of electrons among the $n=0$ and $n=1$ levels and becomes -- in terms of the integral strength and absolute absorption -- the most pronounced line in the transmission spectrum. This simple empirical rule can be justified using the standard linear response theory. Therefore, we focused on this particular excitation in order to find a magneto-optical signature of the surface states in PbSnSe. We applied high magnetic fields to our samples to drive surface electrons close to their quantum limit. At the same time, their fundamental CR mode is expected to move above the reststrahlen band of BaF$_2$. Hence, it should be traceable using infrared magneto-transmission technique.

The latter fact is illustrated in Fig.~\ref{fig:surface_states}.
The LL spectra of bulk and surface electrons, expected for sample C, are shown in panel (a), along with the estimated Fermi energy. In panel (b), we present a false-color plot of magneto-absorbance of sample C overlaid with the expected
energies of bulk and surface inter-LL resonances. The panel (b) shows that the fundamental CR mode of surface electrons should be accompanied by two bulk resonances at nearby energies -- the fundamental bulk CR mode and the lowest bulk interband inter-LL transition. These three excitations are marked by green, cyan and black vertical arrows in Fig.~\ref{fig:surface_states}a. The same color-coding is used for curves in Figs.~\ref{fig:surface_states}b. The dotted parts of the theoretical curves indicate the magnetic fields at which the final-state Landau level is occupied, and therefore, the transition should not appear due to Pauli blocking. Here we considered the bulk Fermi energy to determine the occupation of the surface LLs.

\begin{figure*}
\includegraphics[width=0.75\textwidth]{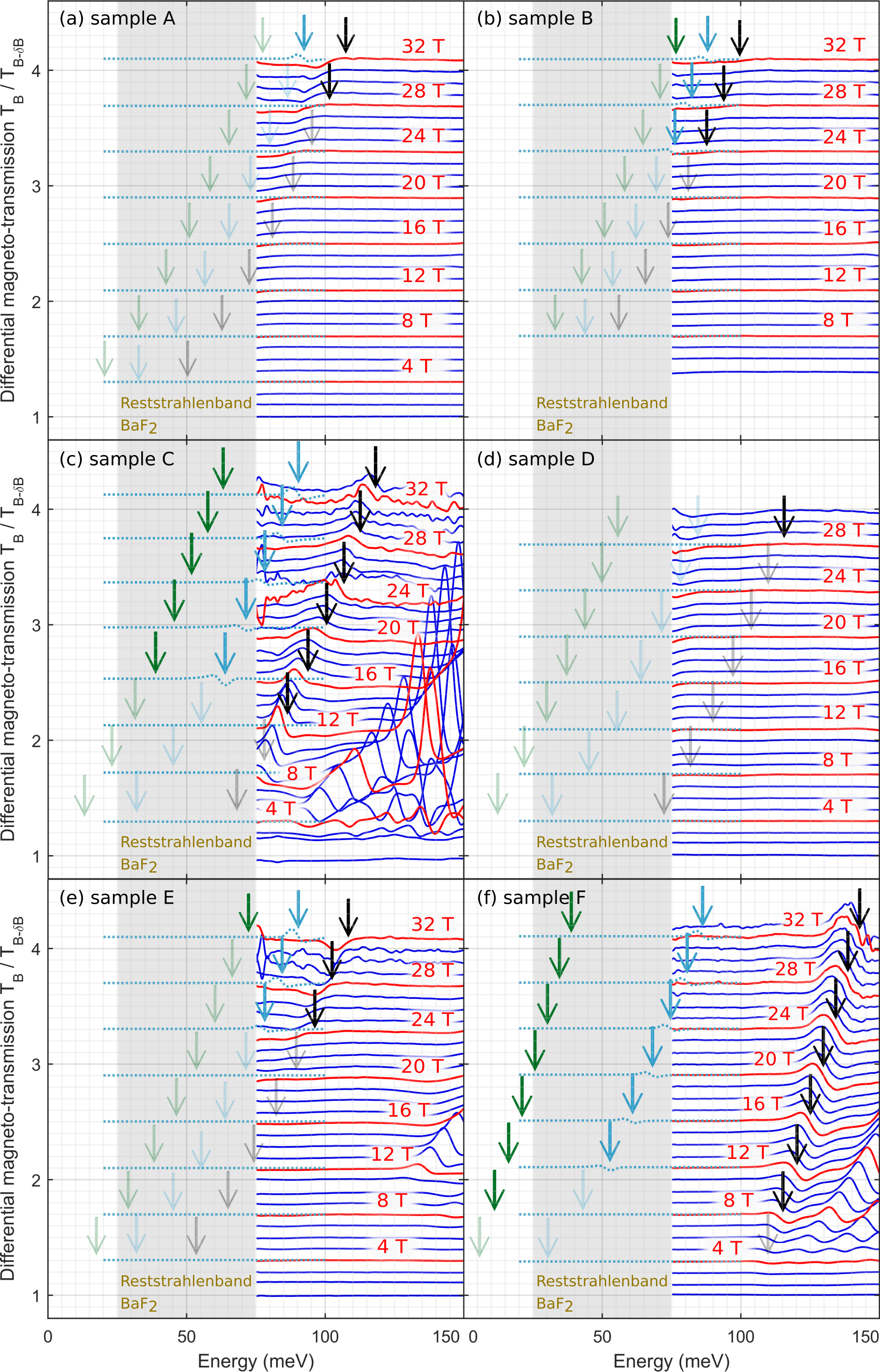}
\caption{Differential magneto-transmission spectra, $T_{B}/T_{B-\delta B}$, plotted for all explored samples A, B, C, D, E, and F in parts (a)-(f), respectively, using the difference $\delta B=1$~T. The spectra are shifted vertically, with the offset scaling linearly with $B$. The energies
of three excitations having the lowest energy in the quantum limit of bulk and surface states are marked by the vertical arrows (cf. Fig.~\ref{fig:surface_states}b): green -- the bulk fundamental CR mode, dark cyan -- the fundamental CR mode of surface electrons and black -- the lowest interband inter-LL transition ($n=0\rightarrow$1). Transitions are expected to emerge
in the spectra (following the occupation effect) when transparent arrows change into solid ones. The dotted cyan line is the theoretically expected response, the differential magneto-transmission, due to the fundamental CR of surface electrons, calculated using the transfer-matrix method, see the text. Neither bulk inter-LL resonances, nor the presence of the reststrahlenband, are included in this model for simplicity.}
\label{fig:exp_theor_bulk_surf}
\end{figure*}

To explore the magneto-optical response in greater details, the differential magneto-transmission spectra, $T_{B}/T_{B-\delta B}$, were analyzed (Fig.~\ref{fig:exp_theor_bulk_surf}a-f). As discussed in the Sec.~III, the impact of a weakly field-dependent Fabry-P\'erot interference pattern is partly suppressed in the differential spectra and we may better follow individual inter-LL resonances. Each transition manifested as a minimum in the relative magneto-transmission, $T_B/T_0$,
is now expected to be seen as a derivative-like feature (a maximum followed by a minimum) in the differential spectra $T_{B}/T_{B-\delta B}$. To facilitate our analysis, the theoretically expected
positions of the three lowest excitations are marked by vertical arrows in Fig.~\ref{fig:exp_theor_bulk_surf}a-f, using the color-coding introduced in Fig.~\ref{fig:surface_states}. The transparent arrows correspond to excitations blocked by the occupation effect. Again, we assume here that the occupation of the surface LLs is given by the bulk Fermi energy. Following these assumptions, the fundamental CR mode of surface electrons should be observable above the reststrahlen band of BaF$_2$ in all explored epilayers, except sample D.

Prior to continuing the analysis of the experimental data, let us discuss the strength of the searched contribution from surface electrons. We performed a simple calculation using the standard transfer-matrix formalism~\cite{Yeh05}. The PbSnSe epilayer was modelled as a dielectric slab with a thickness $d$ (Tab.~\ref{tab:sample_data}) and the refractive index of $n_{\mathrm{PbSnSe}}=5$. The optical response of the surface/interface states were described by the 2D dynamical magneto-conductivity. It comprised a single mode
at $\hbar\omega=E_{\mathrm{2D}}^{n=1}$, with  the  broadening  parameter $\gamma=2$~meV extracted  from  the  bulk  response  and  the integral strength corresponding to the fourfold degeneracy. We assumed that this mode emerges when the bulk Fermi energy drops below the $n=1$ LL of surface electrons.
The substrate was represented by a dielectric slab with a thickness of 500~$\mu$m and a refractive index of $n_{\mathrm{BaF}_2}=1.4$.
The calculated differential spectra (cyan dotted lines) were then plotted along with experimental data in  Fig.~\ref{fig:exp_theor_bulk_surf}. The inflection point in the calculated curves matches the energy of the fundamental CR mode of surface electrons, marked by vertical cyan arrows. When the fundamental CR mode is Pauli blocked, the arrows are transparent. Due to the interference effect, the calculated amplitude of the surface CR mode varies with the thickness of the epilayers. Nevertheless,
the theoretically expected signal clearly exceeds the noise level for all explored samples. For the sake of completeness, let us note that the
strength of the CR mode is much lower as compared to free-standing graphene~\cite{NedoliukNatureNano19} -- the difference in the velocity parameters reduces the strength roughly by a factor of four, further reduction occurs via the dielectric effect, due to PbSnSe bulk slab and BaF$_2$ substrate.

From the experimental viewpoint, the differential spectra measured on samples C and F reveal no  signatures attributable to the signal from the surface states. For other samples, some features develop just above the reststrahlen band of BaF$_2$. Nevertheless, their position, amplitude and shape do not correspond to the expectations for surface electrons. The relatively pronounced minimum observed
for samples A and E (around $\hbar\omega\approx 100$~meV at 30~T) is most likely due to bulk interband inter-LL excitations ($n=0 \rightarrow 1$). Their position seems to be slightly overestimated
in our bulk data modelling (black vertical arrows). The shape -- a minimum instead of derivative-like profile -- may suggest the presence of some additional weaker excitations. However, there is no solid argument connecting them with the surface states.
Below we discuss three different scenarios that allow us to explain why no clear signatures of surface electrons were found in the magneto-optical response: (i) an occupation effect, (ii) disorder and (iii) too simplistic theoretical modelling.

(i) The occupation of surface states, or in other words, the distribution of electrons among the surface LLs, may considerably differ from the one expected from the position of the bulk Fermi energy. The band bending close to the surface or interface of bulk crystals is a well-established phenomenon, often discussed in the context of topological (crystalline) insulators, and other narrow-gap semiconductors~\cite{BansalPRL12,BrahlekSSC15,FrantzeskakisPRX17}. This effect is governed by standard rules of electrostatics and may lead to the appearance of both depletion or accumulation of bulk electrons nearby the surface/interface. The latter possibility could explain the lack of signatures of surface states in our data. Nevertheless, for our samples with relatively large electron doping, and consequently, large density of positively charged defects, a relatively narrow depletion layer and only a weak band-bending effect are expected.

(ii) The topological protection makes surface states robust
against the opening of a band gap, but they still remain sensitive to
disorder which is ubiquitous in a random alloy (note that Sn atoms do
not replace Pb atoms in any periodic or regular manner).
In magneto-optical experiments on 2D systems, the strength of disorder is directly reflected by the widths of inter-LL excitations. In our rule-of-thumb model for the fundamental CR mode of surface electrons, we considered the characteristic broadening equal to that of bulk states. In this way, we might have neglected additional sources of disorder, such as the surface/interface roughness or surface point defects. In fact, our transfer-matrix simulations suggest that an increase of $\gamma$ by a factor of 3-5 brings the calculated magneto-optical signal of the surface states down to the experimental noise level. Disorder in the surface states thus may represent a plausible explanation of the apparently missing signal from the surface electrons.

(iii) The accuracy of the existing theoretical models represents another possible explanation. Clearly, there is no apparent reason to question their general validity~\cite{LiuPHYSREVB2013,HsiehNATCOMMUN2012}, notably, when they are convincingly corroborated by various experimental studies~\cite{DziawaNATMAT2012}. At the same time, however, it is not obvious to what extent these models account for the surface states
quantitatively; strictly speaking, the concept of band structure does
not even apply to a random alloy.
Cushioning such fundamental doubts, the dispersion of the surface states does not need to have its Dirac point in the mid-gap position (reflecting the weak, but present electron-hole asymmetry at the $L$ point). This would effectively correspond to a higher Fermi level for the surface states, thus pushing the quantum limit to larger magnetic fields and making the fundamental CR mode of surface electrons hardly observable in our experiments. In addition, one may expect a certain deviation from the linearity of the conical band on the surface, which is not included in the simplest model~\cite{LiuPHYSREVB2013}. The fundamental CR mode of surface electrons could thus have a lower energy and stay hidden within the reststrahlen band of the BaF$_2$ substrate
even in the highest magnetic fields applied in our experiments. Hybridization of surface and bulk states is another possible effect. Sticking strictly to the standard theoretical model~\cite{LiuPHYSREVB2013}, the bulk and surface states are truly eigenstates of a semi-infinite crystal and they do not mix with each other. At the same time, the discrete surface LLs coexist -- apart from the interval in between zero-mode levels -- with the quasi-continuum of bulk electronic states and one may speculate about mechanisms which mediate their coupling. The hybridization of surface and bulk state may represent an explanation -- alternative to plain disorder -- for spectrally broad features observed above the reststrahlenband of BaF$_2$ in the differential spectra
of the A,B,D and E samples.

\section{Conclusions}

A series of PbSnSe topological crystalline insulator samples has been explored using the infrared spectroscopy technique in high magnetic fields. We show that the fairly rich response observed, comprising a sequence of transitions between bulk Landau levels, can be successfully explained using a simple model of massive Dirac electrons, expanded to include the effects of the band inversion. Despite expectations, and perhaps surprisingly for PbSnSe epilayers with a relatively high electronic quality, we do not find any clear signature of symmetry-protected surface states which are the most salient features of topological crystalline insulators and which could -- similar to topological insulators -- give rise to appealing magneto-optical effects such as universal Faraday and Kerr rotation~\cite{TsePRL10,MaciejkoPRL10,OkadaNC16,ShaoNL17,DziomNC17}. We discuss several scenarios which may account for this: the band-bending effect,  surface disorder and the hybridization between bulk and surface electronic states.

\begin{acknowledgments}
We acknowledge discussions with R.~Buczko, P. E. de Faria Jr., Y.~Fuseya, S.~Tchoumakov, M.~O.~Goerbig and M. Potemski.
The work has been supported by the EC Graphene Flagship project, by the ANR COLECTOR project (ANR-17-CE30-0032), by the French-Czech exchange programme Barrande of MSMT (No. 8J18FR013) and Campus France (No. 40701ZF). XRD and EDS analysis was performed in MGML (www.mgml.eu), which is supported within the program of Czech Research Infrastructures (project no. LM2018096)
We acknowledge the support of the LNCMI-CNRS in Grenoble, a member of the European Magnetic Field Laboratory (EMFL).
\end{acknowledgments}

\bibliography{apstemplateNEW}

\end{document}